\documentclass[conference]{IEEEtran}
\IEEEoverridecommandlockouts

\usepackage{cite}
\usepackage{adjustbox}
\usepackage{amsmath,amssymb,amsfonts}
\usepackage{algorithmic}
\usepackage{graphicx}
\usepackage{textcomp}
\usepackage{xcolor}
\usepackage{enumitem}
\usepackage{multirow}
\usepackage{ragged2e}
\usepackage{url}
\def\BibTeX{{\rm B\kern-.05em{\sc i\kern-.025em b}\kern-.08em
    T\kern-.1667em\lower.7ex\hbox{E}\kern-.125emX}}
    
\begin{document}

\title{Generative Goal Modeling}

\author{\IEEEauthorblockN{Ateeq Sharfuddin}
\IEEEauthorblockA{\textit{School of Computer Science} \\
\textit{Carnegie Mellon University}\\
Pittsburgh, Pennsylvania, United States\\
asharfud@cs.cmu.edu}
\and
\IEEEauthorblockN{Travis Breaux}
\IEEEauthorblockA{\textit{School of Computer Science} \\
\textit{Carnegie Mellon University} \\
Pittsburgh, Pennsylvania, United States\\
breaux@cs.cmu.edu}\\
 \\
}

\maketitle

\begin{abstract}
In software engineering, requirements may be acquired from stakeholders through elicitation methods, such as interviews, observational studies, and focus groups. When supporting acquisition from interviews, business analysts must review transcripts to identify and document requirements. Goal modeling is a popular technique for representing early stakeholder requirements as it lends itself to various analyses, including refinement to map high-level goals into software operations, and conflict and obstacle analysis. In this paper, we describe an approach to use textual entailment to reliably extract goals from interview transcripts and to construct goal models. The approach has been evaluated on 15 interview transcripts across 29 application domains. The findings show that GPT-4o can reliably extract goals from interview transcripts, matching 62.0\% of goals acquired by humans from the same transcripts, and that GPT-4o can trace goals to originating text in the transcript with 98.7\% accuracy. In addition, when evaluated by human annotators, GPT-4o generates goal model refinement relationships among extracted goals with 72.2\% accuracy.

\end{abstract}

\begin{IEEEkeywords}
requirement, elicitation, interview, goal modeling
\end{IEEEkeywords}

\section{Introduction}
\label{section:introduction}
Large language models (LLMs) can perform many natural language processing (NLP) tasks including text summarization, translation, and question-answering with just a few demonstrations and no fine-tuning. Beyond NLP, LLMs offer new opportunities to re-engage hard domain-specific problems. In software engineering, requirements elicitation is one such problem, and a primary technique used to elicit requirements is interviews, where business analysts meet stakeholders to ask questions about the organization, the domain, and existing problems. After conducting an interview, the analyst reviews the interview transcript and identifies requirements. Interviewing is a ``soft skill" that requires experience to be proficient. Subjective information must often be interpreted, different interviewees may respond differently to the same question, and the effectiveness of an interview depends on the appropriateness of the questions.  Furthermore, analysts may either miss requirements in the transcript or introduce requirements based on their own knowledge but not traceable back to the transcript. In this paper, in order to support interviewer eliciting states to be achieved, maintained, or avoided, we describe an approach that applies textual entailment in LLMs to not only extract requirements from interview transcripts but to also identify cases where an interviewer may have missed or introduced requirements. Specifically, we discuss a method to extract requirements traceable to the source text, how the LLM-extracted requirements compare to interviewer-extracted requirements, and the accuracy of the refinement relationships determined by the LLM.

\begin{figure}[t]
\centering
\includegraphics[scale=.32]{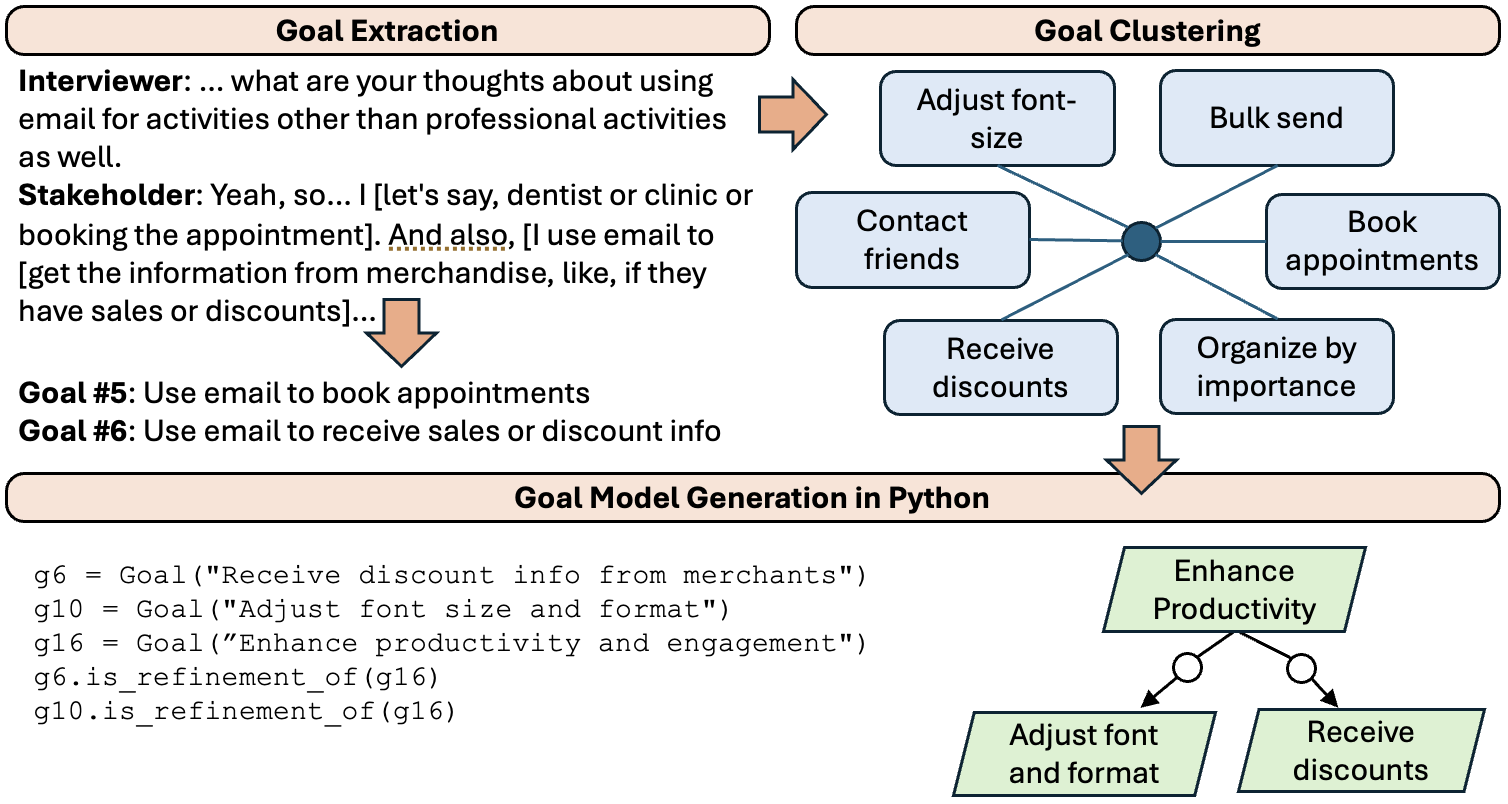}
\caption{Overview of Goal Model Construction Method}
\label{fig:overview}
\end{figure}

The method overview is presented in Figure~\ref{fig:overview}, in which an interview transcript is used to (1) extract goals from interviewer and stakeholder speech while maintaining traceability to the originating phrases in the speech turns. Because LLMs are non-deterministic when sampling tokens, the goal extraction step is performed 10 times to increase the likelihood that the extracted goals will represent any random sampling of the model during this step. While repeated sampling increases goal coverage for each transcript, it also yields duplicate goals. Therefore, (2) the extracted goals are next grouped using agglomerative clustering and cosine similarity. After this step, we randomly sample one goal from each cluster to produce the final list of extracted goals for each transcript. Lastly, (3) the goal model, including refinement relationships, is inferred over the extracted goals, as well as a few implied goals not mentioned by the stakeholder, but inferred from hypothetical refinement relationships that group similar goals into shared categories.

We first review the relevant background in Section~\ref{section:background}. In Section~\ref{section:approach} we detail our overall approach, including how we assembled the transcript dataset, how we prompt for goal extraction, goal clustering, and how we generate goal models. We discuss our evaluation in Section~\ref{section:evaluation}.
Our results are presented in Section~\ref{section:results}. In Section~\ref{section:discussion} we review our findings and go over specific examples. We discuss threats to validity in Section~\ref{section:threats}. Finally, we conclude and consider future research directions.

\section{Background \& Related Work}
\label{section:background}
In this section, we now review related work.

Recently, large language models (LLMs) have reached state-of-the-art performance for information extraction and question-answering tasks. Unlike supervised fine-tuning, in which models are specifically trained to predict a target from an input, LLMs can be instruction-tuned~\cite{OWJ+22} to follow human instructions, which has exhibited added performance gains on natural language benchmarks for reading comprehension, machine translation, natural language inference, and closed-book question answering~\cite{WBZ+22}. 

There are a number of approaches that improve an LLM's reasoning abilities. Chain-of-Thought (CoT) is one approach, where a sufficiently large language model generates a series of intermediate reasoning steps, similar to one's own thought process~\cite{WWS+22}. Similarly, Program-Aided Language (PAL) models generate programs as intermediate reasoning steps, with the final answer being offloaded to a runtime environment, such as Python, for interpretation~\cite{GMZ+23}. Plan-and-Solve Prompting (PSP) is another reasoning approach, where the model devises a plan of subtasks first, and carries out each subtask~\cite{WXL+23}. We adapt these in our approach to generate and refine goals.

Goal-oriented requirements engineering uses goals for requirements elicitation, negotiation, analysis, documentation, and evolution~\cite{Lam09}. In goal-oriented requirements engineering, a \textit{goal model} consists of a refinement graph illustrating how higher-level goals to be achieved, maintained, or avoided by the system are refined into lower-level goals and, conversely, how lower-level goals contribute to higher-level goals~\cite{Lam09}. A goal is a prescriptive statement of intent that the system should satisfy through the cooperation of agents. We describe how to use textual entailment to reliably extract goals from interview transcripts.

Recent research has investigated LLMs for requirements elicitation. Atei et al. (2024) discuss an approach involving LLMs to generate a diverse set of agents simulating users with a wide variety of viewpoints for product design. These agents are then interviewed with a preselected set of questions to describe their product experience, after which an LLM is used to analyze the interview transcripts to identify latent needs~\cite{ACG+24}. Ferrari et al. (2024) discuss generating UML sequence diagrams from requirements documents~\cite{FAA+24}. Chen et al. (2023) discuss producing goal models in Goal-oriented Requirement Language (GRL) for two case studies using GPT-4~\cite{CCH+23}. In a similar theme, Siddeshwar et al. (2024) presented early research using LLMs to generate goal models in GRL from user stories in existing literature~\cite{SSM+24, SSM+24-2}. G{\"o}rer et al. (2023 and 2024) presented approaches utilizing LLMs to generate interview scripts to educate and train interviewers~\cite{GA+23, GA+24}. Wei at al. (2024) utilize LLMs to elicit sub-features from AppStore application features~\cite{WCL+2024}. To the best of our knowledge, our approach is among the first to investigate using textual entailment to extract and refine goals from interview transcripts.

\section{Method \& Approach}
\label{section:approach}

The method shown in Figure~\ref{fig:overview} relies on recent advances in textual entailment enabled by large language models (LLMs)~\cite{BMR+20}, including the use of code generation to represent mathematical structures~\cite{GMZ+23}, such as graphs, which are otherwise difficult for text-to-text models to reason over~\cite{MZA+22}. In this method, we use a multi-modal model GPT-4o that is pre-trained in audio, video, text, and code to map natural language instructions and interview transcript excerpts into code representations of goal models. For the experiments reported in this paper, we use the OpenAI model \textit{gpt-4o-2024-08-06} with \textit{temperature} $= 1.0$, unless otherwise noted.

We now describe the transcript dataset, followed by each of the steps in the method overview. The method evaluation is described separately in Section~\ref{section:evaluation}.

\subsection{Transcript Dataset}

The open-ended interview transcript dataset consists of 34 transcripts recorded by student interviewers over a diverse range of interview topics. We invited students who completed a course assignment on requirements elicitation in a graduate-level requirements engineering class to submit their assignment for research purposes after the completion of the course. The research protocol, including consent process, are actively monitored by an Institutional Review Board (IRB) to protect human subjects.

Prior to the assignment, students received training in requirements elicitation and were assigned to interview another student (the stakeholder) on a topic known to that stakeholder. For training in requirements elicitation, all students attended a one-hour lecture on interviewing and relevant information processing theory from cognitive psychology, including scripts~\cite{SA13}, episodic memory and semantic memory~\cite{Tul83} and how people recall complex events~\cite{Bar88}. In addition, students watched two online videos, entitled ``Tips for Requirements Elicitation"  and ``Common Mistakes in Requirements Elicitation Interviews"  that were produced as part of an empirical study in teaching requirements elicitation~\cite{BZF+19}.

The interview topics were selected using stratified sampling. Before conducting the interviews, in a free-listing exercise~\cite{Bre02}, each student submitted up to 20 different ways that they use the Web, including any mobile applications that can also be accessed from the desktop. This yielded 860 application uses from 79 students that were then categorized into 65 unique topics. Next, each student was randomly assigned one topic from their list, which yielded a topic assignment distribution in which a topic with a higher frequency of mentions across the student responses is more likely to be assigned to a student. Using this assignment method, students were assigned to a total of 29 out of 65 topics, as several topics had one mention and thus would have a low probability of assignment. When students were interviewed, they would be asked to describe their needs as users of a web application in completing one or more tasks in their assigned topic to ensure that interviewees have familiarity with the interview topic. The stakeholder has the web application running on a screen in front but does not share the screen with the interviewer: The stakeholder narrates steps taken to perform various tasks while using the application. The stakeholder narrates what is on the screen, how information is organized on the screen, what controls are available, and what controls the stakeholder chooses to interact with and how. In addition, the stakeholder describes the goals the stakeholder is trying to achieve by performing these actions. The interviewer familiarity with the topic was not controlled and allowed to vary.

Among the original 79 students enrolled in the course, 5 students dropped the course, and 34 students consented to participate, yielding a response rate of 46\%. Participants were compensated with a \$25 Amazon Gift Card for sharing their coursework. Following the IRB-approved research protocol, transcripts were de-identified by removing personally identifiable information prior to use for research purposes. No other changes were made to the transcript text. As part of the assignment, after the interview, the interviewer reviewed the transcript, and identified and created lists of goals and actions that the stakeholder described.

\begin{figure}[t]
\centering
\includegraphics[scale=.40]{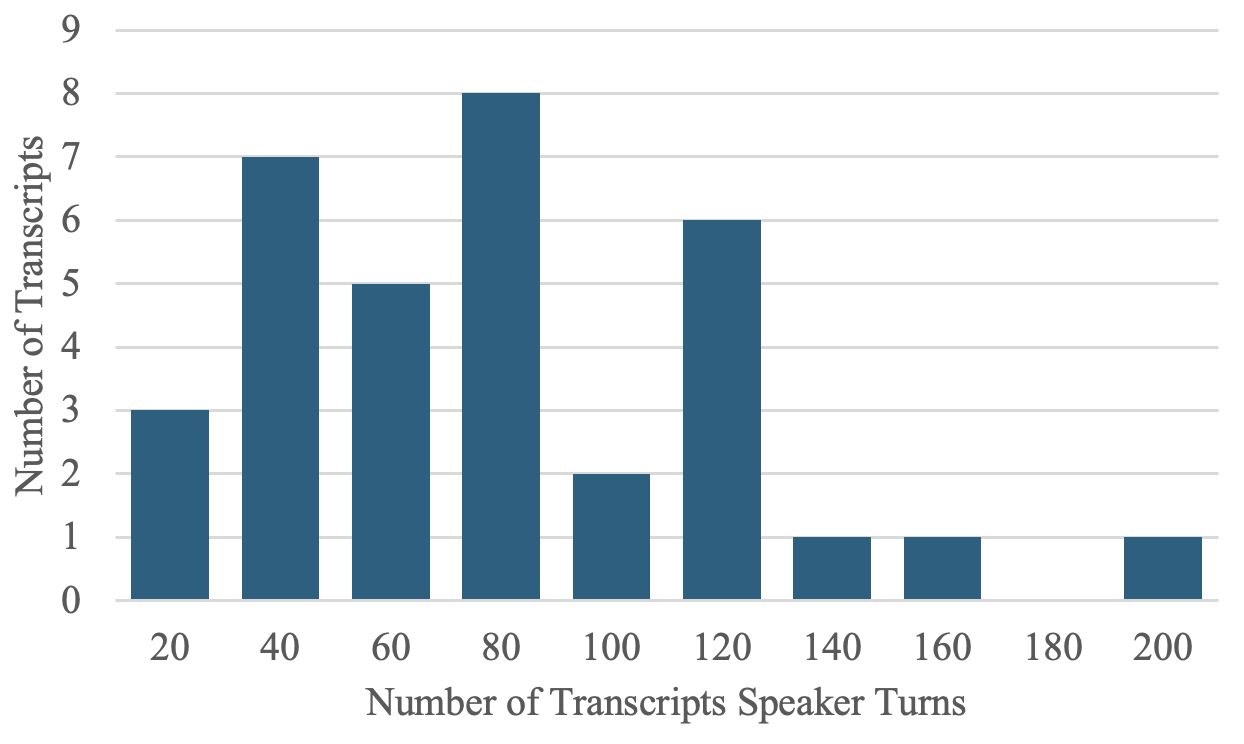}
\caption{Distribution of Transcript Lengths in Speaker Turns}
\label{fig:tran-dist}
\end{figure}

Figure~\ref{fig:tran-dist} shows the distribution of speaker turns over the 34-transcript dataset. The distribution shows that about 80\% of transcripts were between 21-120 turns. The 29 transcript topics cover a wide range of apps, including grocery ordering and other online shopping, investing, job hunting, online education, navigation, photos, podcasts, programming, travel planning, video communications,  instant messaging and e-mail, question-answer forums, news,  online games, weather, social networking, and music and video streaming, among others.

\subsection{Goal Extraction}

During an idealized requirements elicitation interview, one might assume that an interviewer primarily raises questions and confirms their understanding of stakeholder responses, while the stakeholder responds to interviewer questions. Moreover, one might assume that requirements are generally uncovered from stakeholder responses. However, in our experience, interviews are more dynamic and collaborative: Interviewers can pose hypothetical scenarios or introduce requirements directly as questions, while stakeholders may ask their own clarification questions, after which the interviewer may respond by revealing requirements-related details of their own that the stakeholder may subsequently adopt or confirm. Therefore, to capture requirements from interviews, one must recognize that requirements can span multiple speaker turns and may not be limited to stakeholder turns.

With the large context windows supported by commercial LLMs (e.g., 128,000 tokens or about 96,000 words), one approach could be to submit an entire transcript to the model. One concern, however, is that transformer-based models appear to lose attention in the middle of large input contexts~\cite{LLH+24} (see Section~\ref{section:background}), called a lead and/or trailing bias~\cite{RSC+24}, which could result in goals being missed in the extracted goal list. Therefore, an alternate approach is to use an extraction window that limits the number of speaker turns included in the input context. Because a goal may trace back to more than one speaker turn in the transcript history, an extraction window should be incremented by a separate number of turns that is less than the window length. This decision yields a moving window, which may have the unwanted outcome of extracting duplicate goals, because the goal-yielding speech overlaps more than one window. Based on these two approaches, we conducted a comparative evaluation of prompting using a window of length equal to the whole transcript, versus a moving window with length equal to four turns that moves by increments of two turns.

Independent of the window size, goal extraction proceeds in two steps: first to extract the goals and second to trace the extracted goals back to the transcript. Given the transcript excerpt of window-length, GPT-4o is prompted to extract goals from the excerpt using the prompt template and instruction shown in Figure~\ref{fig:prompt1}. The instruction quotes the goal definition from van Lamsweerde's Requirements Engineering textbook~\cite{Lam09}, and requests that goals not reference specific products or services to yield system-independent goals. We experimented by replacing the italicized sentence in Figure~\ref{fig:prompt1} with ``Do not include references to applications, products or services in the goal statement.'' to test the effect of generalization on the goal extraction and goal model generation quality. The response format uses JavaScript Output Notation (JSON) to aid in parsing the LLM response. We experimented with writing goals in KAOS format, e.g., Achieve[EmailsFilteredByTags] in which the goal type of achieve, maintain, or avoid is explicit and the goal state is written in past tense. However, due to the added effort required to shape the output through demonstrations, we chose to postpone this translation to future work.

\begin{figure}[ht]
\begin{tabular}{p{8.5cm}}
\fontsize{7}{5}{\texttt{Read the following interview transcript excerpt and respond with any goals that the speaker expresses. A goal describes a prescriptive statement of intent that the system should satisfy through the cooperating of its agents. \textit{Write goals in general terms and do not include references to applications, products or services in the goal statement.} Only write goals that can be traced to specific phrases in the speech. Respond with the goals in a JSON list of strings.}} \\
\fontsize{7}{5}{\texttt{\{transcript\_excerpt\}}} \\
\fontsize{7}{5}{\texttt{Response:}}
\end{tabular}
\caption{Prompt 1 to Extract Goals}
\label{fig:prompt1}
\end{figure}

Second, we combine the transcript excerpt with the extracted goals in a second prompt (see Figure~\ref{fig:prompt2}) to extract the candidate source text from which the goals were plausibly generated. The authors manually inspected every extracted source text and determined whether the matched text was evidence of the given goal. We investigated other methods to identify the source text, including semantic similarity metrics, but we found that prompting GPT-4o was superior and yielded a high tracing accuracy, which we report in Section~\ref{section:results}.

\begin{figure}[ht]
\begin{tabular}{p{8.5cm}}
\fontsize{7}{5}{\texttt{Read the following the goals in JSON format and identify the substrings in the interview transcript excerpt from which the goals were generated. Respond in JSON using the format \{"goal": "goal statement", "phrases": ["phrase1", "phrase2"]\}}} \\
\fontsize{7}{5}{\texttt{\{transcript\_excerpt\}}} \\
\fontsize{7}{5}{\texttt{\{generated\_goals\}}} \\
\fontsize{7}{5}{\texttt{Response: }}
\end{tabular}
\caption{Prompt 2 to Trace Extracted Goals to Speech}
\label{fig:prompt2}
\end{figure}

Finally, we use role-based prompting~\cite{WPQ+24} using same system message with Prompts \#1 and \#2: ``You are a business analyst collecting requirements for a software application. Your job is to review interview transcripts between an interviewer who is a business analyst and a stakeholder who is a prospective user of the application. The interviewer will ask the stakeholder questions to identify their requirements for the application.''

\begin{figure}[t]
\centering
\includegraphics[scale=.4]{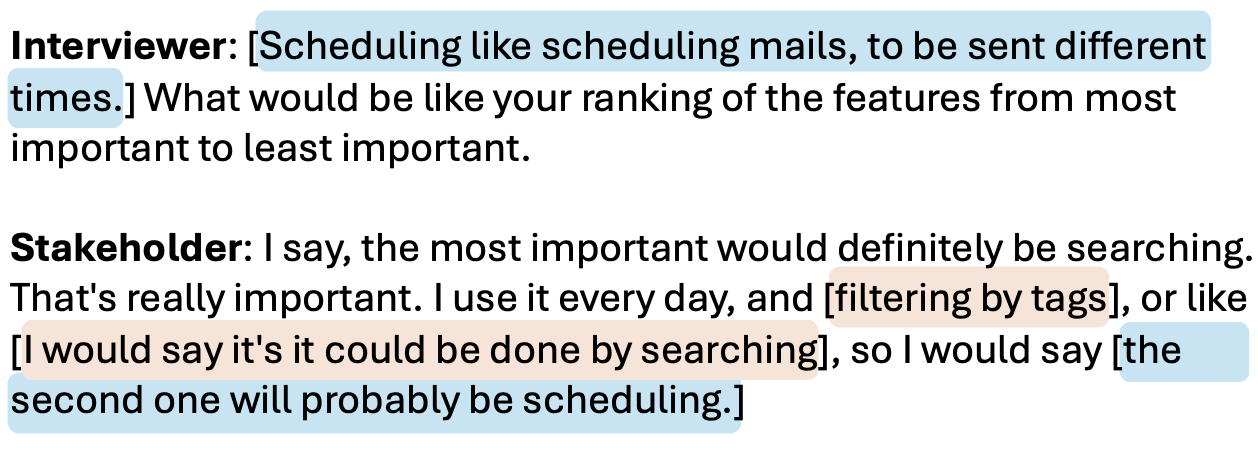}
\caption{Example Traces for Goal Extraction}
\label{fig:example-trace}
\end{figure}

Figure~\ref{fig:example-trace} presents an example output trace from Prompt \#2 applied to an excerpt in Transcript \#4 that traces the two goals: (1) ``Implement filtering by tags feature'' (orange) and (2) ``Implement scheduling capability for sending emails'' (blue). The source text from the transcript is identified by brackets with the color highlight added for presentation, herein. The filtering goal illustrates a goal identified by the stakeholder in response to a question. The scheduling goal illustrates how a goal can be traced across multiple turns, e.g., interviewer and stakeholder speech. In this example, the interviewer is re-introducing a solution previously mentioned by the stakeholder and requesting prioritization.

With the extracted goals per turn and source text, we next discuss how we cluster the extracted goals to minimize overlaps and remove duplicates.

\subsection{Goal Clustering}

Large language models commonly use the softmax function in the last layer of the underlying neural network to convert a vector of real numbers into a probability distribution from which tokens are sampled when generating a text sequence. Temperature scaling~\cite{GPS+17} is a technique to ``soften'' the softmax function: when temperature $T \rightarrow \infty$ the uncertainty increases, which can lead to different generations from the same input. At $T = 1$, the original probabilities are recovered, whereas at $T \rightarrow 0$, the probability collapses to point mass, in which case the model generates the same output. Repeated sampling with temperature scaling can lead to non-deterministic output. Because each token probability depends on previously generated tokens, repeated sampling can lead to generating different outputs.

In our experience, the non-determinism in output leads to variations in extracted goals: some goals are missed in one generation, but captured in another generation. To address this issue, we re-sample the goal extraction step multiple times. However, repeated sampling can also lead to duplicate goals. To resolve duplicates, we apply agglomerative clustering to group goals from all samples by computing the Euclidean distance between their text embeddings obtained using a sentence transformer. Next, we randomly sample one goal from each cluster, which are then carried forward into the goal model construction step.

We use Python v3.11.11 and SciKit Learn v1.6.1 with the AgglomerativeClustering class and parameters \textit{metric=euclidean}, \textit{linkage=ward}, and \textit{distance\_threshold=1.5}, in addition to the \textit{paraphrase-MiniLM-L6-v2} sentence transformer model for clustering.

Below, we present four clusters consisting of goals extracted by GPT-4o from Transcript \#4. Clusters \#1 and \#2 illustrate a mixed-goal cluster with goals that describe different stakeholder needs under one topic, whereas clusters \#3 and \#4 are more typical and narrowly describe a single goal with multiple paraphrases. As we discuss in Section~\ref{section:discussion}, around 80\% of clusters observed are single-goal clusters, and we discuss method improvements in that section, such as mixed-goal cluster sampling.

\begin{itemize}
    \item[\textbf{1:}] [`Check emails daily', `Search emails effectively', `Filter emails', `Read unread emails and mark important ones', `Simplify the process of reading multiple emails', `Search emails daily', `Tag emails for later action', 'Forward important emails to others', `Mark emails to revisit later', `Unsending or editing sent emails instead of sending a new one', `Search for emails efficiently', `Ensure important emails are not missed by checking the spam folder', `Keep inbox manageable by deleting unnecessary emails']
    \item[\textbf{2:}] [`Open the browser to access email', `Upload multimedia to the cloud and share links instead of attachments']
    \item[\textbf{3:}] [`Book appointments', `Book appointments', `Book appointments', `Book appointments', `Book appointments', `Book appointments with services like clinics', `Schedule appointments such as dental or clinic visits']
    \item[\textbf{4:}] [`Get email from recruiters', `Receive emails from recruiters', `Receive emails from recruiters daily', `Get emails from recruiters', `Receive emails from recruiters', `Check for new information from recruiters', `Receive emails from recruiters', `Receive emails from recruiters', `Receive emails from recruiters', `Receive emails from recruiters', `Stay informed about job opportunities by checking emails from recruiters']
\end{itemize}

We next discuss how the randomly sampled goals are used to generate goal models.

\subsection{Goal Model Generation}

One approach to build a goal model would be to query a reliable goal modeler as to whether a refinement relation $r$ exists between two goals $g_{1}$, $g_{2}$ in the model $M$. Given a set of $n$ goals, there are $n(n-1)/2$ possible refinement relations between goals in a goal model represented by a complete, directed acyclic goal graph. For example, a goal model with 30 goals could have up to 435 refinement relations. In our experience, goal models do not maximize the number of refinement edges in the graph, in which case most inquiries $r(g_{1}, g_{2}) \in M$ will be untrue. As goal models increase in size, this can become computationally expensive for little benefit.

Another approach inspired by how humans build goal models is, given a randomly sorted list of goals $L$, iterate over the goals $g_i \in L$, choosing a goal $g_{1}$ while looking for a second goal $g_{2}$ with maximal probability of a refinement relationship $r(g_{1}, g_{2}) \in M$. As relations are discovered, attention shifts to the goals remaining in the list without declared relations. In addition, high-level goals, which can be observed by finding goals in the goal graph without incoming edges, become obvious starting points for finding where to define a new relation. Under both approaches, it is possible that goals in the goal list have a low probability of refinement relations with any other goal, i.e., the goals are all relatively independent, for example, they may be sibling goals or they may plausibly be under different parent goals that are not present in the list. In this case, the goal modeler may infer \textit{implied goals} from the goal list that describe latent high-level goals or purposes for other goals in the list. To realize this approach, we designed a prompt with the following four components based on prior work in prompting design. The prompt instruction shown in Figure~\ref{fig:prompt3} includes the following elements:

1) A code completion task to represent the goal graph in Python based on a pre-defined Python class with member variables and functions to leverage program-aided language (PAL) model-based reasoning~\cite{GMZ+23}. We chose this approach, because many instruction-tuned models are bimodal, i.e., they are trained on natural language text and code, and generating graph-like structures in text produces syntax errors~\cite{MZA+22}, which requires additional effort to correct and which may be uncorrectable, if the errors are ambiguous. In contrast, Python is a language learned during pre-training that can be compiled to check for syntax errors and the compiled code and symbol table can be used for computing transformations over the graph. In addition, PAL-based reasoning appears to be better calibrated than Chain-of-Thought (CoT) reasoning~\cite{KRM+24}.

2) Steps that guide the model in planning the code, before generating a solution. This strategy is similar to chain-of-thought (CoT)~\cite{WBZ+22} and plan-and-solve prompting (PSP)~\cite{WXL+23}, which has shown improved performance on knowledge reasoning tasks. Within these steps, we instruct the model to generate an overall app description and to enumerate the definitions of each goal including what is and is not included in the meaning. Similar to CoT and PSP, these two outputs provide context for generating refinement relations.

3) Guidance for introducing and defining implied goals, before generating code that declares refinement relations between the implied and explicit goals.

4) Guidance on how to format the output using enclosing \fontsize{9}{5}{\texttt{```python}} and \fontsize{9}{5}{\texttt{```}} tags that can be matched to extract the code. Because the generated code can include print statements to standard output that may interfere with a later execution environment, we include an instruction to prohibit generating print statements.

5) A generic one-shot Python expression to demonstrate the different ways to create refinement relations between goals and to collect the implied goals. Demonstrations have been shown to improve performance in early large language models~\cite{BMR+20}. Our preliminary experiments found that one only needs to demonstrate specific function calls, whereas demonstrating entire programs can lead the model to generate code that builds on expressions from the demonstrations, which is incorrect. 

6) A Goal class definition that includes member variables and functions that the model may assign to or call. This portion of the prompt serves as a code stub to be completed by the model and introduces discipline about what shape the function calls will take and what subsequent calls are most probably given the restricted definition.

Within the prompt template, the \fontsize{9}{5}{\texttt{\{goal declarations\}}} slot is filled with declarations in the format: \fontsize{9}{5}{\texttt{g4 = Goal("See percentage of chapter read")}}\fontsize{9}{5}{. Prior to instantiating the template, each goal is assigned a unique number once that is then reused across prompts.

\begin{figure}[ht]
\begin{tabular}{p{8.5cm}}
\fontsize{7}{5}{\texttt{Read the following Python code that describes an initial goal model and complete the code using the following function calls. Before completing the code, perform the following steps: 1) read all of the goals and describe what the software application is and does; 2) for each goal, describe what the goal means in the context of the application description, including what is and is not intended by the goal description; 3) identify up to two or three *implied* goals that were not included in the original goal list and that add missing context to the original goals; 4) extend the code by adding your implied goals; and 5) complete the code by reviewing all of the goals and deciding which goals are refined by other goals. Implied goals include high-level goals that group related refinements together, and explain what actions the low-level goals seek to achieve. For each implied goal that you create, add the goal to the list of implied goals. Include your justification for each function call in comments. Do not write code to print the goals. When responding, include the python code between the start ```python and end ``` tags.}} \\
\fontsize{7}{5}{\texttt{1. X.is\_refined\_by.append(Y) - when the goal X is satisfied by the goal Y and has the refinement goal Y}} \\
\fontsize{7}{5}{\texttt{2. Y.is\_refinement\_of(X) - when the goal Y satisfies the goal X and is a refinement of goal X}} \\
\fontsize{7}{5}{\texttt{3. implied\_goals.append(X) - to collect the implied goals that you created}} \\
\fontsize{7}{5}{\texttt{class Goal: }} \\
\fontsize{7}{5}{\texttt{\quad def \_\_init\_\_(self, text):}} \\
\fontsize{7}{5}{\texttt{\quad \quad self.text = text}} \\
\fontsize{7}{5}{\texttt{\quad \quad self.is\_refined\_by = []}} \\
\fontsize{7}{5}{\texttt{\quad \quad def is\_refinement\_of(self, goal):}} \\
\fontsize{7}{5}{\texttt{\quad \quad \quad goal.is\_refined\_by.append(self)}} \\
\fontsize{7}{5}{\texttt{implied\_goals = []}} \\
\fontsize{7}{5}{\texttt{\{goal declarations\}}}
\end{tabular}
\caption{Prompt 3 to Generate Goal Model}
\label{fig:prompt3}
\end{figure}

We used the following system message with Prompt \#3: ``You are a business analyst building a goal model of stakeholder goals for a software application. Goal models are directed, acyclic graphs in which edges trace from high-level goals to low-level goals through refinement relationships. High-level goals describe what states stakeholders want to achieve, maintain or avoid in the system. Low-level goals describe *how* the system will satisfy high-level goals, tend to be more specific and describe how the system will operate. High-level goals describe *why* the system aims to satisfy low-level goals, tend to be more generic and describe what the stakeholder aims to accomplish independent of a specific software application.''

For each goal-set extracted from a transcript, we generate 10 goal models using repeated sampling and Prompt \#3. During each generation, if a model does not compile or the model contains cycles, we re-generate that goal model. We evaluate the generated goal models using the method described in Section~\ref{section:evaluation}.

Figures~\ref{fig:prompt3a}~and~\ref{fig:prompt3b} illustrate excerpts from the GPT-4o output of Prompt \#3, in which Chain-of-Thought (CoT) is routinely produced in response to the planning instruction and as part of the comments to code. Figure~\ref{fig:generated-model} shows one of the goal graphs from the excerpt of Prompt \#3. The generated CoT serves to ground the task in domain-relevant vocabulary and should not be interpreted as an explanation, sound justification, or source of validity or evaluation.

\begin{figure}[ht]
\begin{tabular}{p{8.5cm}}

\fontsize{7}{5}{\texttt{**1. Software Application Description:**
The software application appears to be an email management software designed to assist users in managing their emails effectively. It focuses on organizing incoming emails, summarizing information, scheduling communication, and notification management...}} \\
\fontsize{7}{5}{\texttt{}} \\
\fontsize{7}{5}{\texttt{**2. Goals Context:**}} \\
\fontsize{7}{5}{\texttt{- **g0:** "Receive emails from recruiters" implies that the user wants to manage job-related communications effectively. It does not necessarily cover email handling beyond the receipt of emails from recruiters.}} \\
\fontsize{7}{5}{\texttt{}} \\
\fontsize{7}{5}{\texttt{- **g1:** "Receive notifications from the school" suggests a focus on educational communications. It does not cover reading or taking actions on these emails...}} \\
\fontsize{7}{5}{\texttt{}} \\
\fontsize{7}{5}{\texttt{**3. Implied Goals:**}} \\
\fontsize{7}{5}{\texttt{- "Organize and classify incoming emails": This group offers a high-level goal for sorting and prioritizing emails, which aligns with several goals like g7 and g11.}} \\
\fontsize{7}{5}{\texttt{- "Enhance email productivity and engagement": Provides context for g6, g9, and g10 by encompassing actions and tools to improve email communication.}} \\
\fontsize{7}{5}{\texttt{- "Ensure email accessibility": A broader context for g13 and g4, emphasizing access and status updates on incoming emails.}} \\
\end{tabular}
\caption{Example Goal Modeling Plan Generation for Prompt 3}
\label{fig:prompt3a}
\end{figure}

\begin{figure}[ht]
\begin{tabular}{p{8.5cm}}
\fontsize{7}{5}{\texttt{g0 = Goal("Receive emails from recruiters")}} \\
\fontsize{7}{5}{\texttt{g1 = Goal("Receive notifications from the school")}} \\
\fontsize{7}{5}{\texttt{g2 = Goal("Use email to book appointments")}} \\
\fontsize{7}{5}{\texttt{g3 = Goal("Contact with friends")}} \\
\fontsize{7}{5}{\texttt{g4 = Goal("Check emails daily")}} \\
\fontsize{7}{5}{\texttt{g8 = Goal("Understand key points and summarize received emails")}} \\
\fontsize{7}{5}{\texttt{g9 = Goal("Summarize emails using an AI assistant")}} \\
\fontsize{7}{5}{\texttt{g14 = Goal("Check email for important information daily")}} \\
\fontsize{7}{5}{\texttt{}} \\
\fontsize{7}{5}{\texttt{\# Implied Goals}} \\
\fontsize{7}{5}{\texttt{g15 = Goal("Organize and classify incoming emails")}} \\
\fontsize{7}{5}{\texttt{implied\_goals.append(g15)}} \\
\fontsize{7}{5}{\texttt{\# Refinements}} \\
\fontsize{7}{5}{\texttt{g8.is\_refinement\_of(g15)  \# Understanding key points and summarizing are ways to organize.}} \\
\fontsize{7}{5}{\texttt{g9.is\_refinement\_of(g8)   \# Summarizing using AI is a refinement of understanding and summarizing.}} \\
\fontsize{7}{5}{\texttt{g4.is\_refined\_by.append(g14)  \# Checking for important information specifies the daily checking.}} \\

\end{tabular}
\caption{Example Goal Model as Generated Code from Prompt 3}
\label{fig:prompt3b}
\end{figure}

\begin{figure}[t]
\centering
\includegraphics[scale=.07]{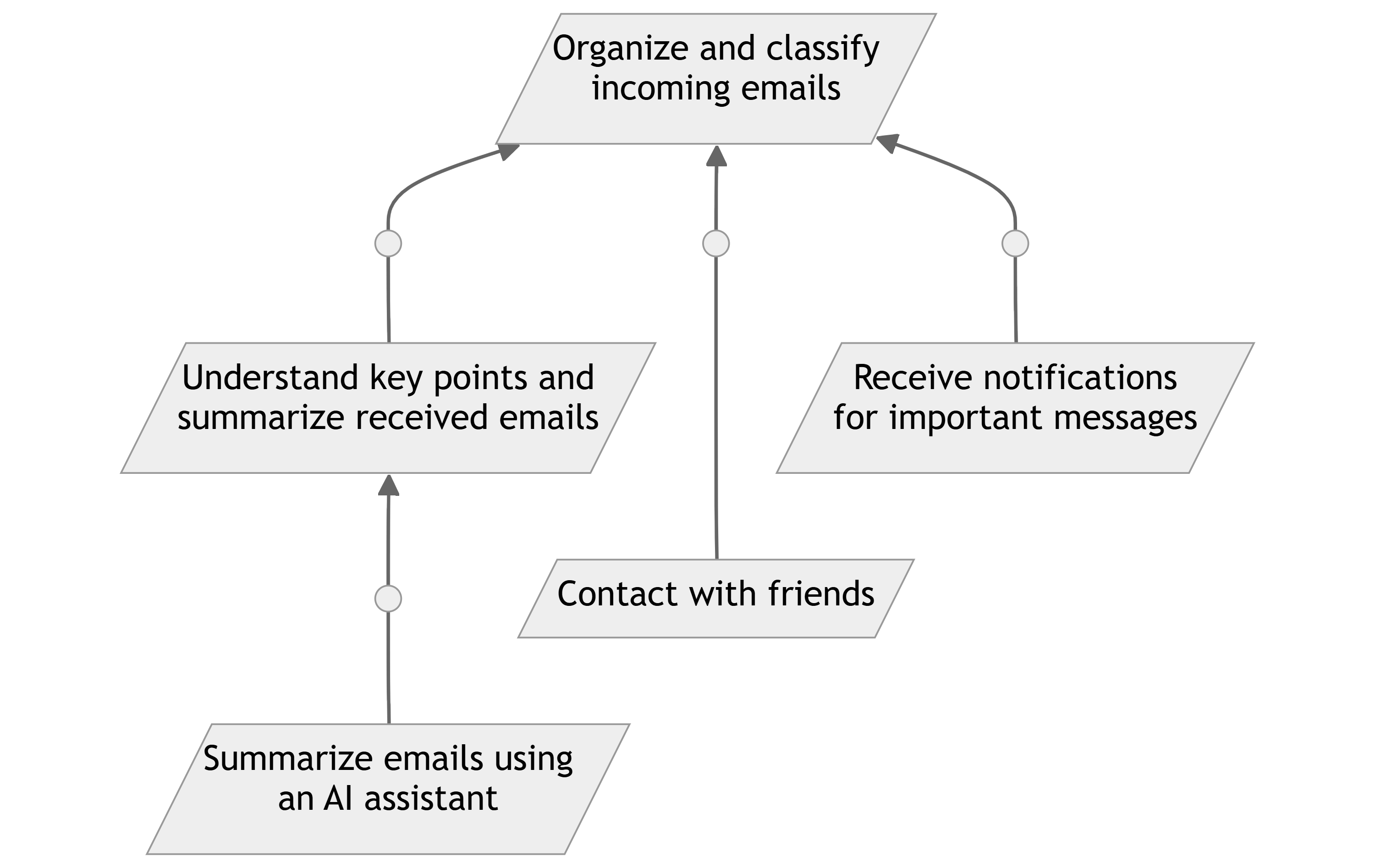}
\caption{One of the goal graphs from Generated Code from Prompt 3}
\label{fig:generated-model}
\end{figure}

\section{Evaluation}
\label{section:evaluation}

We assess the accuracy and reliability of goal extraction and goal generation.

\subsection{Goal Extraction}

\begin{flushleft}
    \justifying
    \textbf{RQ1:} How well do extracted goals trace back to source text?
\end{flushleft}

Traceability between extracted goals and source text in the interview provides credibility and context to each goal. To answer RQ1, we apply Prompt \#2 (see Figure~\ref{fig:prompt2}) to the transcript to find all possible string candidates for the human- and GPT-4o-extracted goals. The tracing prompt does not limit the number of source phrases that support a goal, nor does it distinguish between interviewer or stakeholder speech when attributing a goal to a source phrase. Correctness is measured by the first and second author manually verifying the trace between the goal and source text. In Section~\ref{section:results} we report the correctness as a ratio of correct traces to total traces, the number of goals exclusively traced to interviewer speech and stakeholder speech, the number of goals that trace to multiple speaker turns and the average distance between speaker turns. We also report the distribution of phrase locations across the transcript, normalized by transcript length. Large language models have a documented leading and trailing bias that can lead to low attention to information contained in the middle of the context~\cite{LLH+24,RSC+24}. The presence of this bias would be represented by a higher frequency of goals being extracted from the beginning and end of the transcript.

\begin{flushleft}
    \justifying
    \textbf{RQ2:} How do LLM-extracted goals compare to human-extracted goals?
\end{flushleft}

We compare the interviewer-extracted and GPT-4o-extracted goals in two steps. First, we label the goals in each set with one of four category labels: \textit{high-level goal}, which means the goal describes a general state to be satisfied and has a low probability of being a sub-goal; \textit{low-level goal}, which means the goal describes a specific state to be satisfied and has a high probability of being a sub-goal; \textit{soft-goal}, which means the goal describes a quality to be minimized or maximized by the system; and \textit{goal regress}, which means the statement describes a state outside of the system scope. Second, the first and second authors label the LLM-extracted goals with identifiers that uniquely identify human-extracted goals with a close approximation of the same meaning. The authors developed heuristics that they reused throughout the mapping process to consistently address edge cases. Third, we use the source text traces identified in response to RQ1 to detect goals that derive from the same source texts. The reported findings include the proportion of overlapping goals by each category, and proportion of manually identified overlaps that also share overlapping source phrases in the transcript.

\subsection{Goal Model Generation}
\begin{flushleft}
    \justifying
    \textbf{RQ3:} What is the accuracy of refinement relationships in LLM-generated models?
\end{flushleft}

To answer RQ3, we extracted refinement relationships between a parent and child goal from the acyclic goal model graph for all goal models generated by the model. We distinguish between those relationships \textit{declared} in the generated goal model, and any relationships that do not appear in the model we call undeclared. First, we remove any cyclic graphs to yield the graph set $G$ and then compute the set of vertices $v$ and edges $e$ for each graph $(v, e) \in G$, including the edges in the transitive closure of $e$, which we obtain by the function $T(e)$. Next, we computed the undeclared edges by subtracting the edges in the closure $T(e)$ from the edges in the complete graph $C(v)$ for all vertices $v$. Finally, we sampled an equal number of declared and undeclared edges from $T(e)$ and $C(v) - T(e)$, respectively. The sampled edges were sorted into random order in a labeling task, in which each author independently labeled each edge as a true refinement relation, if vertex $p$ is refined by or refines vertex $c$ in $(p, c) \in e$; or as false refinement, if the vertex $p$ is not refined by or does not refine vertex $c$. The authors kept notes concerning their observations during labeling, including whether an edge described two equivalent goal statements. The authors labeled 50 edges independently, which yielded a Kappa $\kappa = 0.288$, which is fair agreement~\cite{LK77}. Next, they reviewed their disagreements and developed four rules to clarify labels as follows: 

\begin{itemize}
    \item For goals $g_{1}$ and $g_{2}$, if $g_{1}$ describes an activity in broader terms than $g_{2}$, then $g_{1}$ is refined by $g_{2}$;
    \item If $g_{2}$ is a step in the a process described by $g_{1}$, then $g_{1}$ is refined by $g_{2}$;
    \item If $g_{1}$ describes an environmental action by an agent that the system supports through $g_{2}$, then $g_{1}$ is refined by $g_{2}$; and
    \item If $g_{1}$ describes the same action as $g_{2}$, with or without an additional action, including a pre- or post-condition, then $g_{1}$ is equivalent to $g_{1}$. 
\end{itemize}

After developing these rules, the authors coded a second sample of 50 refinement relations to yield an updated Kappa $\kappa = 0.711$, which is a substantial level of above chance agreement~\cite{LK77}. Next, the first author labeled the remaining refinement relations to obtain a dataset of 550 relations. Finally, the labeled ground truth is used to evaluate the model-generated refinement relations by computing the number of true positives (TP), true negatives (TN), false positives (FP) and false negatives (FN). Accuracy is reported using the formula: (TP + TN) / (TP + TN + FP + FN).

\section{Results}
\label{section:results}

The goal extraction method was applied to 34 transcripts, which cover a total of 2,615 turns. The average number of turns per transcript is 76.9 turns with the median number of 70 turns. Table~\ref{table:rq1} presents summary statistics from the goal extraction step: for GPT-4o, the \textbf{whole} column describes the extraction results from using the whole transcript, the \textbf{moving} columns describe extraction using the moving window with an instruction to respond with only general goal descriptions (-G) and without this additional instruction (-S), and \textbf{Human / moving} is the tracing method applied to the human extracted goals. Each extraction was performed 10 times and the table reflects the average statistics: \textit{total phrases} is the number of phrases traced to the total number of extracted goals; \textit{unmatched phrases} is the proportion of total phrases that did not match strings within the transcript (i.e., hallucinated phrases); \textit{trace correctness} is the ratio of matched phrases to total phrases; \textit{interviewer goals} is the number of goals that trace to matching speech in the interviewer's turns; \textit{stakeholder goals} is the number of goals that trace to matching speech in the stakeholder's turns; and \textit{multi-turn goals} are goals that span across multiple turns.

In Table~\ref{table:rq1}, we observe that GPT-4o in general produced on average 105 more goals than what the humans reported. During our inspection of the human-authored goals, we observed goals that did not derive from the transcript. Rather, they were likely introduced by the interviewer using their own knowledge, which is a similar observation to Debnath et al. (2021)~\cite{DSF21,FSD22}. The inability to trace human-authored goals to the transcript, if indeed they were created solely by the interviewer, may explain the lower 83.3\% trace correctness, despite using the moving window for tracing. 

\begin{table}[h!]
\caption{Statistics for Goal Extraction Task}
\label{table:rq1}

\centering
\begin{tabular}{ |l|r|r|r|r| } 
 \hline
 \multirow{2}{*}{\textbf{Statistics}} & \multicolumn{3}{c|}{\textbf{GPT-4}} & \textbf{Human} \\
 \cline{2-4}
& \textbf{whole} & \textbf{moving-G} & \textbf{moving-S} & \textbf{moving} \\ 
 \hline
 Total phrases & 672.2 & 604.7 & 535.5 & 595.2 \\
 \hline
 Unmatched phrases & 109.4 & 34.8 & 31.3 & 99.6 \\
 \hline
 Trace correctness & 83.7\% & 94.2\% & 94.1\% & 83.3\%\\
 \hline
 Interviewer goals & 6.5 & 17.8 & 13.2 & 16.1\\
 \hline
 Stakeholder goals & 339.9 & 335.5 & 265.3 & 232.7\\
 \hline
 Multi-turn goals & 6.5 & 13.5 & 8.7 & 13.5\\
 \hline
 Total goals & 346.4 & 353.3 & 278.5 & 248.8 \\
 \hline
\end{tabular}
\end{table}



The RQ1 asks ``How well do extracted goals trace back to source text?'' To answer RQ1, the authors manually reviewed all the transcript traces tied to these goals and found that GPT-4o incorrectly produced two goals, resulting in an overall accuracy of 98.7\%. We discuss these two cases in Section~\ref{section:discussion}. 

The RQ2 asks ``How do LLM-extracted goals compare to human-extracted goals?'' To answer RQ2, the authors manually mapped the GPT-4o extracted goals to the human-authored goals. We observed that 173 human-authored goals mapped to GPT-4o-generated goals. GPT-4o also produced additional 30 goals that traced to the transcript and that the interviewers had missed. In addition, there were 76 human-authored goals that did not map to any GPT-4o-generated goal. Overall, the goals generated by GPT-4o matched 62.0\% of the goals generated by the interviewers. We discuss examples of the differences further in Section~\ref{section:discussion}. 

The RQ3 asks ``What is the accuracy of refinement relationships in LLM-generated models?'' To answer RQ3, the second author randomly sampled from refinement relations generated by GPT-4o and relations not generated by this model, and the first author labeled a randomized shuffle of the sample for whether they are true or false refinement relations. Among these results, we observed that GPT-4o-generated 227 true positives, 170 true negatives, 73 false positives and 80 false negatives for an overall accuracy of 72.2\%. We elaborate with specific examples in Section~\ref{section:discussion}.

\section{Discussion}
We now discuss our observations in agglomerative clustering and goal extraction.

\label{section:discussion}
\subsection{Agglomerative clustering}
Agglomerative clustering works by iteratively merging pairs of clusters based on a distance metric, such as Euclidean distance. The benefit of this method is that one need not specify the number of clusters in advance, e.g., as in k-means clustering. Consequently, this technique does not guarantee that all goals in a cluster are semantically equivalent, even if they are more similar than goals outside the cluster. Below, we present three clusters from a Transcript \#32 describing how the stakeholder prepares for a programming interview. In cluster \#1, each of the goals refer to efficiency in using the keyboard to write code. This includes goals that describe efficiency in the absence of a mouse, and a goal to avoid using UI-heavy tools. In cluster \#2 obtained from goals extracted from the same transcript, however, the cluster covers a wider range of topics, including unrelated goals. Finally, cluster \#3 includes goals to support multiple programming languages. In our inspection of the 22 clusters created from goals for this transcript, we observed that 20 clusters are more like cluster \#1 and \#3, which describe near-similar goals, and only two clusters are like cluster \#2 with a mix of dissimilar goals. We informally observe this phenomena across multiple transcript clusters and attribute it to agglomerative method, which terminates using a pre-defined distance threshold when the remaining clusters are too dissimilar.
\begin{enumerate}
    \item[\textbf{1:}] [`Ensure efficiency by keeping hands on the keyboard while coding.', `Keep hands on keyboard to improve efficiency', `Improve efficiency with keyboard shortcuts', `Map keyboard shortcuts for specific functions', `Maintain a productive workflow using keyboard shortcuts', `Achieve efficiency without using a mouse', `Remember commands and shortcuts without referring to documentation', `Eliminate reliance on UI-heavy tools', `Use keyboard shortcuts for coding tasks']
    \item[\textbf{2:}] [`Push code to version control', `Keep config file to make the system just like local customization', 'Edit text-based files', `Switch between files quickly', `Switch between files easily']
    \item[\textbf{3:}] [`Fit tools for any kind of programming language', `Fit editor for any kind of language', `Support any kind of programming language', `Fit text editor for any programming language', `Allow customization for any kind of language', `Handle different programming languages', `Support any programming language']
\end{enumerate}

The consequence of mixed-goal clusters is that random sampling from clusters like cluster \#2 will lead to under-sampling dissimilar goals. For example, if we randomly choose ``Push code to version control'' from cluster 3, then we miss goal ``Edit text-based files'', which is notably dissimilar. One solution would be to detect these dissimilar clusters and increase the sampling within the cluster to include dissimilar goals, for example, up to some threshold of dissimilarity. To address this issue, we propose to extend the method by checking intra-cluster similarity to find the 1-2 mixed goal clusters and then to select goals using a distance threshold and metric, such as cosine similarity.

\subsection{GPT-4o Goals Missed by the Interviewers}
We have a number of goals that GPT-4o extracted that the interviewers missed. The authors manually reviewed and traced all these goals to the texts in the transcripts. With the exception of two goals, we believe the rest of the goals were missed simply due to human oversight. We highlight the two goals generated by GPT-4o for which we did not find evidence.

\begin{enumerate}
\item \texttt{Save and read posts when offline.} In Transcript \#28 the stakeholder discusses an \textit{obstacle} instead of a goal, where one cannot perform these activities as a consequence of being offline.
\item \texttt{Express appreciation for personalized playlists.} In Transcript \#5 the stakeholder voices really liking a particular feature during the interview, and this is not a goal.
\end{enumerate}

\subsection{Interviewer Goals Missed by GPT-4o}

As mentioned in Section \ref{section:results}, there are two goals solely authored by the interviewer that cannot be traced to the transcripts. In the first case (Transcript \#5), the interviewer listed ``search songs by title" as a goal, whereas the text describes searching for songs based on emotions the stakeholder feels or based on keywords but not by title. In the second case (Transcript \#9), the interviewer introduced ``training and implementing AI tools to analyze user preferences..." as a goal, whereas the stakeholder generically describes preferring one platform that uses AI over another platform that does not. We consider these as cases where the interviewers used their own knowledge to author these goals. Although these goals were not declared by the stakeholders in the interview, they were potential goals to be achieved in the systems. 

\subsection{Effect of Transcript Excerpt Size}

Figure~\ref{fig:goal-dist} presents the distribution of the extracted goals across all 34 transcripts, binned by the transcript partition for 10 partitions per transcript. Longer transcripts with 191 turns would average 19.1 turns per partition, whereas a shorter transcript with 27 turns would average 2.7 turns per partition. In the figure, when the extraction window covers the entire transcript, we observe a leading bias in which goal extraction yields more goals within the first through third tenths of the transcript and the fewest goals in the last tenth of the transcript. By comparison, when using a moving window of four turns, which increments by two turns each move, then goal extraction is more evenly distributed throughout the transcript for both GPT-4o-generated and human-authored goals that were traced to the windowed excerpt.

\begin{figure}[t]
\centering
\includegraphics[scale=.4]{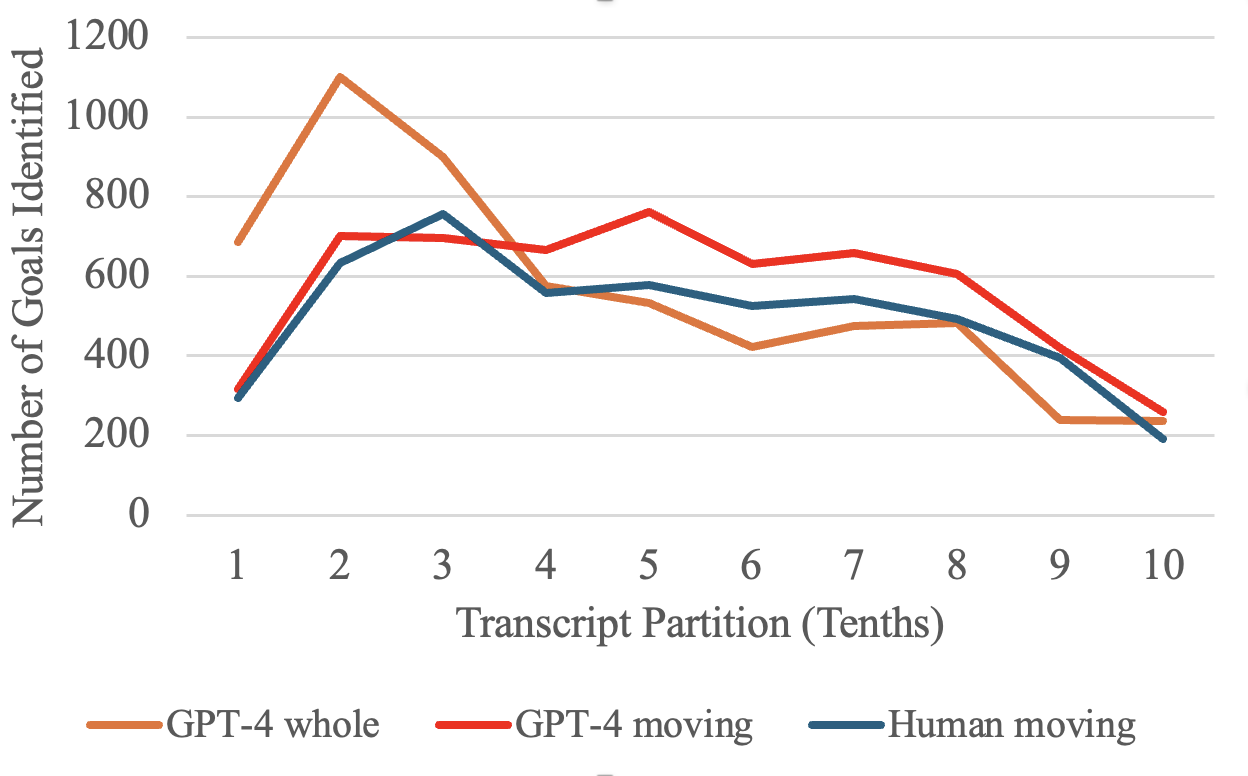}
\caption{Frequency of Goals Extracted over Transcript, 10 Partitions}
\label{fig:goal-dist}
\end{figure}

\subsection{Analysis of Goal Refinement Relations}

We conducted an analysis of goal refinement relations by examining relations that contain cycles and by measuring additional dimensions of the generated refinement relations in comparison to the ground truth dataset. The ground truth dataset is based on 15 transcripts, which were used to generate 10 goal models per goal set extracted from each of the 15 transcripts. This yielded 150 generated goal models, which contain a total 2,372 refinement relations. We analyzed the edges between goals for directionality to detect any symmetries, i.e., where one model defines a relation between two goals $g_{1},g_{2}$ as a ``refines'' relation, whereas another model defines the relation as a ``refined by'' relation. This analysis identified 47 symmetric relations distributed over 13/15 models. In some cases, the symmetries were between near synonymous goals: ``Receive job notifications unless they complicate the notification tab'' and ``Receive job postings instantly when they are posted.'' In others, the goals were different but topically related: ``Consider employee reviews, current work tasks, and tech stack in job postings'' and ``Navigate information about the tech stack more easily.''

We grouped the GPT-4o-generated refinement relations that did not match our assigned labels into one of the following categories.
\begin{enumerate}
\item GPT-4o generates symmetric refinement relations across two models from the same prompt, whereas we define the relation as a refinement: e.g., ``Receive job postings instantly when they are posted'' and ``Receive job notifications unless they complicate the notification tab'' are not symmetric. There were 23 cases.
\item GPT-4o does not generate a relationship, whereas we define a refinement relation: e.g., ``Receive frequent notifications about rain in real time'' and ``Provide timely and accurate weather information.'' Our review indicated that all these instances were specific goals refining broader goals. There were 80 such cases.

\item GPT-4o generates a refinement relation, whereas we expect no relationship: e.g., ``Book flights early for better prices'' and ``Make sure flight details work for the schedule.'' We noticed that these cases were complementary goals. There were 65 cases.

\item GPT-4o generates a refinement relation between $g1$. which is \textit{refined by} $g2$, whereas we consider $g1$ to be a \textit{refinement} of $g2$: e.g., ``Share documents instantly with teammates'' and ``Accessible and integrated communication system.'' In all these cases, the specific goal should refine the broader goal. There were 35 cases.
\item GPT-4o generates a symmetric refinement relation, whereas we claim no relation exists: e.g., ``Communicate with friends and family'' and ``Backup and restore messages when changing phones.'' The goals are not related. There were 6 cases.

\item GPT-4o generates a refinement relation between $g1$, which is \textit{refined by} $g2$, whereas we claim no relation exists: e.g., ``Enhance search functionality on the platform'' and ``Filter potential connections based on employment status.'' The goals are not related. There were 2 cases.

\item GPT-4o does not generate a relationship, whereas we define a symmetric relation: ``Receive job notifications'' and ``Receive job recommendations.'' We see these goals as equivalent. There was 1 such case.

\end{enumerate}

The dataset, including the transcripts, notebooks, and scripts are all available under the GitHub repository\\ \url{https://github.com/cmu-relab/goalgeneration}.

\section{Threats to Validity}
\label{section:threats}

We now discuss threats to validity.

\textit{Construct validity} is the correctness of operational measures used to collect data, build theory and report findings from the data~\cite{Yin03}, and the extent to which an observed measurement fits a theoretical construct~\cite{SCC02}. Axel van Lamsweerde defines a goal as ``a prescriptive statement of intent that the system should satisfy through the cooperating of its agents''~\cite{Lam09}. In the KAOS goal modeling framework, goals are defined in terms amenable to real-time temporal logic, as a state or event to be achieved, maintained, or avoided by the system~\cite{DLF93}. For example, the goal ``Implement filtering by tags feature'' extracted by GPT-4o from Transcript \#4 could be written as the achievement goal Achieve[EmailsFilteredByTags] in KAOS. Yu defines goals as statements about ``\textit{why} a system is needed,'' specifically highlighting organizational goals~\cite{Yu97}. In requirements engineering, actions that are controllable or observable by the machine can be described by \textit{specifications}~\cite{Jac95}. In addition, Zave and Jackson describe goals that are not directly impacted by the machine as \textit{goal regress}~\cite{ZJ97}. In addition to goal statements, goal refinement relationships can exist in a AND-refinement relationships to indicate that the satisfaction of two or more goals is necessary to satisfy a parent goal.

In our method, we instruct GPT-4o using the goal definition introduced by van Lamsweerde (see Prompt \#1), quoted above~\cite{Lam09}. This yields a range of high and low-level action descriptions that begin with a verb. We did not evaluate the extracted goals for whether they were observable or controllable, nor did we check for whether goal regress was occurring, which may be a threat to construct validity. We did evaluate whether the extracted goals were similar to human-authored goals in response to RQ2. Finally, our definition of refinement relationship excludes AND-refinements and only covers OR-refinements.

\textit{Internal validity} is the extent to which measured variables cause observable effects in the data~\cite{Yin03}. In this study, we introduced multiple points of measurement to check that assumptions were correct between the stages of goal extraction and goal model generation. This includes checking that extracted goals were valid descriptions of interviewer and stakeholder speech in response to RQ1. In addition, we evaluated the goal refinement relations, including relations generated and not generated by GPT-4, as well as, relations generated between explicit and implicit goals in response to RQ3.

\textit{External validity} determines the scope of environmental phenomena or domain boundaries to which the theory and findings generalize~\cite{Yin03}. We identify two threats to external validity. (1) The transcripts describe student interviewers. Although the student interviewers were taking a graduate-level requirements engineering class and received lecture, training, and videos to prepare them to interview stakeholders, this assignment might have been a student interviewer's first time conducting requirements elicitation. Furthermore, the student interviewers could have blended interviewing concepts from other communications courses that introduce interviewing techniques not directly related to requirements elicitation. The stakeholder responses obtained by professional software engineers with industry experience conducting interviews may differ from those obtained by the student interviewers in this study. Nonetheless, the interview assignments were evaluated against a rubric and graded to the satisfaction of the requirements engineering class instructors. (2) The prompts and prompt outputs (e.g., extracted goals and generated goal models) are limited to GPT-4o. Prior research shows that prompt performance does not transfer across models~\cite{LBM+22}, even across different models sizes of the same family (e.g., 13B versus 175B GPT-3). Transferability is increasingly difficult as models undergo different fine-tuning practices (e.g., instruction-tuning~\cite{WBZ+22}, alignment~\cite{OWJ+22} and function calling~\cite{SDD+23}).

\section{Conclusion}
\label{section:conclusion}
In software engineering, interviews are a popular technique to elicit requirements. Successful interviewing requires experience to direct the dialogue with an aim of creating a specification. In addition, the interviewer can miss requirements stated in the transcript or they can use their own knowledge to introduce requirements that do not trace back to the transcript. We developed an LLM-based approach to extract requirements from interview transcripts and to build goal models that organize requirements into refinement relationships. This includes finding goals explicitly stated in transcripts, in addition to implied goals inferred from the extracted goal list. The evaluation yields a 72.2\% accuracy in finding refinement relationships and an error analysis identifies a number of explanations for the false positives and false negatives. Future work includes new techniques, such as self-reflection, to reduce this error and to introduce additional goal model elements, such as agents, operations, conflicts, and obstacles.

\section*{Acknowledgment}

This research was funded by NSF Awards \#2007298, \#2217572, \#2217573, and \#2317987.

\bibliographystyle{IEEEtran}
\bibliography{references.bib}

\vspace{12pt}

\end{document}